\documentclass[preprint,1p]{elsarticle}					
\usepackage{graphicx}						
\newcommand{\up}{\uparrow}
\newcommand{\dn}{\downarrow}
\journal{Annals of Physics}

\begin{document}
\begin{frontmatter}
\title{Spin dynamics of large-spin fermions in a harmonic trap}
\author{Junjun Xu$^1$}
\author{Tongtong Feng$^1$}
\author{Qiang Gu}
\ead{qgu@ustb.edu.cn}
\address{Department of Physics, University of Science and Technology Beijing, Beijing 100083, China}
\address{$^1$These authors contributed equally to this work.}

\begin{abstract}
Understanding the collective dynamics in a many-body system has been a central task in condensed matter physics. To achieve this task, we develop a Hartree-Fock theory to study the collective oscillations of spinor Fermi system, motivated by recent experiment on spin-9/2 fermions. We observe an oscillation period shoulder for small rotation angles. Different from previous studies, where the shoulder is found connected to the resonance from periodic to running phase, here the system is always in a running phase in the two-body phase space. This shoulder survives even in the many-body oscillations, which could be tested in the experiments. We also show how these collective oscillations evolve from two- to many-body. Our theory provides an alternative way to understand the collective dynamics in large-spin Fermi systems.
\end{abstract}
\end{frontmatter}

\section{Introduction}
The origin of collective behaviours has been a central problem in condensed matter physics, ranging from the spin waves in magnetic materials to the Majorana fermions in topological insulators. Spinor Bose gases have provided a unique system to test these collective properties in the past two decades \cite{Kawaguchi, Stamper}, including the spin dynamics \cite{Pu1, Pu2, Chang, Zhang2, Mahmud}, thermal condensates \cite{Pechkis, He}, and spin textures \cite{Mizushima, Stamper2, Leslie, Kawakami, Choi, Eto}. Among these, one of the central issues is to understand the collective spin-mixing dynamics of such system\cite{Ketterle1, Ketterle2, Chapman}. Such coherent dynamics has been observed\cite{Schmaljohann, Chang1, Kuwamoto}, reflecting the quantum nature of the spinor Bose-Einstein condensate. This kind of spin dynamics is driven by the spin-changing collision\cite{Ho, Ohmi}, and could be significantly affected by the external magnetic field\cite{Pu3, Romano, Zhang, Chang2, Kron, Black}. A resonance-like phenomenon was predicted by calculating the field-dependence of the oscillation amplitude and period\cite{Zhang}, and was confirmed in experiments\cite{Kron, Black}. Moreover, coherent spin-mixing dynamics in an ultracold spinor bosonic mixture was reported very recently\cite{Li}.

Renewed interest in this regard has been strongly stimulated by the progress in the spinor fermionic atoms, for example, $^{6}{\rm Li}$, $^{40}{\rm K}$ and $^{173}{\rm Yb}$, who may carry spin $f=3/2$, $9/2$, and $5/2$, respectively, larger than $1/2$ \cite{Yip, Wu, Wu2, Demler, Ueda, Taie, Gorshkov, Taie2, Rey, Rey2, Fallani, Fallani2, Folling, Ho2, Krauser1, Pu4, Krauser2, Ebling}. Similar to the spinor Bose gases, there exist spin-changing collision channels which could cause spin changing for large-spin Fermi atoms\cite{Krauser1, Pu4, Krauser2, Ebling}. However, since the Fermi gases as a whole cannot be described as a globle macroscopic wave function, there remains an intriguing question whether the spin-mixing process could exhibit coherent oscillation or not. This question was just answered by the Hamburg group with their pioneering experiments \cite{Krauser1,Krauser2}. They carried out experiments on the quantum degenerate Fermi gases of $^{40}{\rm K}$ whose hyperfine ground state has a total spin $f=9/2$ and thus contains ten spin states with magnetic quantum numbers $m_{f}=-9/2,\ldots,9/2$. The system is initially prepared in a coherent state with spins rotated about $x$-axis with angle $\theta$. Long-lived and large-amplitude spin oscillations driven by the $\left|m_f=1/2\right>+\left|m_f=-1/2\right>\leftrightarrow\left|m_f=3/2\right>+\left|m_f=-3/2\right>$ collision channel were observed when the system was quenched to sufficiently low temperature and magnetic field. Besides the oscillation behavior, experiment results indicates that the oscillation damps \cite{Krauser1,Krauser2}. Similar phenomenon was also observed in spinor condensates \cite{Schmaljohann, Chang1}.

How to understand this collective dynamics of a large-spin Fermi system? Compaired to previous studies using quantum Boltzmann equation to explain the experimental data\cite{Krauser2,Ebling}, in this paper, we propose an alternative description of the spin-mixing dynamics based on the Hartree-Fock (HF) approximation \cite{Ashcroft, Xu}. According to \cite{Baym}, these two methods are connected to each other. However, one key advantage of our HF approach is that it gives insight into two-body physics, and is beyond semi-classical approximation. We can also in this method observe these collective oscillations from two- to many-body.

Without loss of generality, we take the fermionic system with hyperfine spin $f=3/2$ as an example. We consider the atoms in a one-dimensional trapping potential. To derive the spin-mixing term, we suppose that each single-particle state is a superposition of all the four spin states. Two physical restrictions have to be satisfied: conservation of total magnetization and the Pauli exclusion principle. Such restrictions ensure that two atoms with hyperfine spins of $m_f=1/2$ and $-1/2$ may change into two atoms with $m_f=3/2$ and $-3/2$, respectively, and vice versa. However, the process of converting two atoms with $m_f=-1/2$ and $-3/2$ into two both with spin $m_f=1/2$ is forbidden.

The paper is organized as follows. First we give our model and develop a time-dependent HF theory for spinor fermions. We then study the oscillation dynamics for two atoms and show how these oscillations emerge as a collective behaviour in a many-body Fermi system. We focus on two different rotation angles and show how the oscillation period behaves for these two cases. At last, we give our summary.

\section{The Model}
We consider a system consists of spinor Fermi gases with hyperfine spin $f=3/2$. The system is initially prepared in a harmonic trap with spin magnetic quantum number $m=\pm1/2$. A radio-frequency pulse is used to rotate the spins about the $x$-axis with angle $\theta$ to create a coherent state. Then the interaction and magnetic field is turn on, which drive the spinor dynamics between $m=\pm1/2$ and $m=\pm3/2$ states. The Hamiltonian of the system is $H=H_0+H_I$, with the non-interaction one
\begin{eqnarray}
H_{0}=-\frac{1}{2}\nabla^{2}+\frac{1}{2}x^{2}-pF_z+qF_z^2,
\end{eqnarray}
where the last two terms account for the linear and quadratic Zeeman effect. Here we have use the dimensionless quantities $x=\tilde{x}/\sqrt{\hbar/(m\omega)}$ with $\tilde{x}$ is the position of the atoms with $\omega$ the trap frequency. $p=g\mu_B B/(\hbar\omega), q=(g\mu_B B)^2/(E_{hf}\hbar\omega)$ are reduced linear and quadratic Zeeman energy with $E_{hf}$ the hyperfine splitting, $B$ the magnetic field, $g$ the Land\'e factor, and $\mu_B$ the Bohr magneton. The two-body interaction $H_I$ can be projected to the total spin space
$|F,m_{F}\rangle$
\begin{eqnarray}
H_{I}=\sum_{F=0}^{2f-1}\sum_{m_{F}=-F}^{F}g_{F}\delta(x-x^{\prime})|F,m_{F}\rangle\langle F,m_{F}|,
\end{eqnarray}
where $g_{F}=\tilde{g}_F/(\hbar\omega\sqrt{\hbar/m\omega})$ is the reduced two-body interaction strength in total spin-$F$ channel. Since we are trying to capture the physics of three-dimensional (3D) system, here $\tilde{g}_F$ is connected to the 3D $s$-wave scattering length $a_F$ by $\tilde{g}_F\approx 2\hbar^2a_F/[ml_\perp^2(1-a_F/l_\perp)]$ \cite{Guan}, where $l_\perp=\sqrt{\hbar/(m\omega_\perp)}$ is the transverse width with $\omega_\perp$ the harmonic frequency in the $y$ and $z$ direction. If $\vert a_F\vert\ll l_\perp$, we have $g_F=2\gamma a_F/(\sqrt{\hbar/(m\omega)})$ with $\gamma=\omega_\perp/\omega$. In an elongated trap $\gamma\gg 1$, we have $a_F\ll g_F/2\sqrt{\hbar/(m\omega)}$. Since we only consider $s$-wave interaction here, so the two-body real space wavefunction is symmetric, thus the spin space wavefunction is antisymmetric, i.e. $F=0,2$. So in this system we have two interaction parameter $g_{0}$ and $g_{2}$. The Hamiltonian $H$ commutes with $F_{z}$ but not $F$. So for a collision process, the magnetic quantum number $m_{F}=m_{1}+m_{2}$ is conserved. However, the total spin $F$ can change between $0$ and $2$.

The wavefunction of the system $\Psi$ is written as a Slater determinant
of single-particle wavefunction
\begin{eqnarray}
\psi_n^{s=\up,\dn}(x)&=&\phi^s_{n,\frac{3}{2}}(x)\vert3/2\rangle+\phi^s_{n,\frac{1}{2}}(x)\vert1/2\rangle\nonumber \\
&&+\phi^s_{n,-\frac{1}{2}}(x)\vert-1/2\rangle+\phi^s_{n,-\frac{3}{2}}(x)\vert-3/2\rangle\nonumber\\
&=&\left[\phi^s_{n,\frac{3}{2}},\phi^s_{n,\frac{1}{2}},\phi^s_{n,-\frac{1}{2}},\phi^s_{n,-\frac{3}{2}}\right]^T,\label{eq:singleparticlewf}
\end{eqnarray}
where $n$ labels the energy level and the spin direction of the state $s=\up,\dn$ (here we label the spin of each atom as $\uparrow$ or $\downarrow$) is projected to the whole spin space with $m=\pm3/2,\pm1/2$. We will see in the following that the spin $s$ of the single-particle state is rotating during the dynamical process, which shows the spinor behaviour of the system. The wavefunction of the system is now readily written as
\begin{eqnarray}
\Psi(\mathbf{x})=\frac{1}{\sqrt{(2n)!}}\left\vert \begin{array}{cccc}
\psi_1^{\uparrow}(x_{1}) & \psi_1^{\uparrow}(x_{2}) & \cdots & \psi_1^{\uparrow}(x_{2n})\\
\psi_1^{\downarrow}(x_{1}) & \psi_1^{\downarrow}(x_{2}) & \cdots & \psi_1^{\downarrow}(x_{2n})\\
\vdots & \vdots & \ddots & \vdots\\
\psi_n^{\uparrow}(x_{1}) & \psi_n^{\uparrow}(x_{2}) & \cdots & \psi_n^{\uparrow}(x_{2n})\\
\psi_n^{\downarrow}(x_{1}) & \psi_n^{\downarrow}(x_{2}) & \cdots & \psi_n^{\downarrow}(x_{2n})
\end{array}\right\vert ,
\end{eqnarray}
where $\mathbf{x}=(x_{1},x_{2},\cdots,x_{2n})$. Note that even though there are four degenerate single-particle states for each energy level $n$, we take account only two of them with $s=\up,\dn$. This is due to our initial state configuration and the system size is $2n$. The energy of the system can be written as
\begin{eqnarray}
E=\int\Psi^{*}(\mathbf{x})(H_{0}+H_{I})\Psi(\mathbf{x})d\mathbf{x}=E_{0}+E_{I},
\end{eqnarray}
where
\begin{eqnarray}
E_{0}&&=\sum_{n,s,m}\int\phi_{n,m}^{s*}\left(-\frac{1}{2}\nabla^{2}+\frac{1}{2}x^{2}-pF_{z}+qF_{z}^{2}\right)\phi^s_{n,m}dx\nonumber \\
&&=\sum_{n,s,m}\int\phi_{n,m}^{s*}\left(-\frac{1}{2}\nabla^{2}+\frac{1}{2}x^{2}\right)\phi^s_{n,m}dx\nonumber\\
&&+\sum_{n,s}\int\left[\frac{9q}{4}\left(\vert\phi^s_{n,\frac{3}{2}}\vert^{2}+\vert\phi^s_{n,-\frac{3}{2}}\vert^{2}\right)dx+\frac{q}{4}\left(\vert\phi^s_{n,\frac{1}{2}}\vert^{2}+\vert\phi^s_{n,-\frac{1}{2}}\vert^{2}\right)\right]dx
\end{eqnarray}
is the non-interaction energy. Hereafter, we left out the $x$ in $\phi^s_{n,m}$ if that does not cause any confusion. We can see the linear Zeeman term makes no contribution to the system due to the symmetry of the $\pm m$ state. The interaction energy is calculated using Clebsch-Gordan coefficients. For example, the spin-mixing term
\begin{eqnarray}
&&\int\left[\langle1/2\vert\phi_{n,\frac{1}{2}}^{s*}(x_{1})\otimes\langle-1/2\vert\phi_{n',-\frac{1}{2}}^{s'*}(x_{2})\right]H_{I}\nonumber\\
&&\times\left[\phi^{s'}_{n',-\frac{3}{2}}(x_{2})\vert-3/2\rangle\otimes\phi^s_{n,\frac{3}{2}}(x_{1})\vert3/2\rangle\right]dx_{1}dx_{2}\nonumber \\
&=&\int\phi_{n,\frac{1}{2}}^{s*}\phi_{n',-\frac{1}{2}}^{s'*}\phi^{s'}_{n',-\frac{3}{2}}\phi^s_{n,\frac{3}{2}}dx\sum_{F=0,2}g_F\langle m_{1}=1/2,m_{2}=-1/2\vert F,m_F=0\rangle\nonumber\\
&&\times\langle F,m_F=0\vert m_{1}=3/2,m_{2}=-3/2\rangle\nonumber \\
&=&(g_{2}-g_{0})/4\int\phi_{n,\frac{1}{2}}^{s*}\phi_{n',-\frac{1}{2}}^{s'*}\phi^{s'}_{n',-\frac{3}{2}}\phi^s_{n,\frac{3}{2}}dx.
\end{eqnarray}
We can see that this spin-mixing process will disappear if $g_{0}=g_{2}$. In our following calculations, we will only focus on the effect of $g_0$ and set $g_2=0$ for simplicity. Now we are ready to get the interaction energy, which is straightforward as
\begin{eqnarray}
E_{I}&=&\sum_{nn'ss'}g_0/2\int dx\left(\vert\phi_{n,\frac{1}{2}}^s\vert^{2}\vert\phi_{n',-\frac{1}{2}}^{s'}\vert^{2}+\vert\phi_{n,\frac{3}{2}}^s\vert^{2}\vert\phi_{n',-\frac{3}{2}}^{s'}\vert^{2}\right.\nonumber\\
&&-\phi_{n,\frac{1}{2}}^{s*}\phi_{n',-\frac{1}{2}}^{s'*}\phi_{n,-\frac{1}{2}}^{s}\phi_{n',\frac{1}{2}}^{s'}-\phi_{n,\frac{3}{2}}^{s*}\phi_{n',-\frac{3}{2}}^{s'*}\phi_{n,-\frac{3}{2}}^{s}\phi_{n',\frac{3}{2}}^{s'}\nonumber\\
&&+\phi_{n,\frac{1}{2}}^{s*}\phi_{n',-\frac{1}{2}}^{s'*}\phi_{n,-\frac{3}{2}}^{s}\phi_{n',\frac{3}{2}}^{s'}+\phi_{n,\frac{3}{2}}^{s*}\phi_{n',-\frac{3}{2}}^{s'*}\phi_{n,-\frac{1}{2}}^{s}\phi_{n',\frac{1}{2}}^{s'}\nonumber\\
&&\left.-\phi_{n,\frac{1}{2}}^{s*}\phi_{n',-\frac{1}{2}}^{s'*}\phi_{n,\frac{3}{2}}^{s}\phi_{n',-\frac{3}{2}}^{s'}-\phi_{n,\frac{3}{2}}^{s*}\phi_{n',-\frac{3}{2}}^{s'*}\phi_{n,\frac{1}{2}}^{s}\phi_{n',-\frac{1}{2}}^{s'}\right).
\end{eqnarray}
The first four terms on the right are the usual HF term and the last four terms account for the spin-mixing between $m=\pm1/2$ and $m=\pm3/2$ states. The time-dependent HF equations are get by the variation of $E-\mu N$to $\phi^s_{n,m}$ and then set $\mu$ to $i\hbar\partial_{t}$, which is
\begin{eqnarray}
\left(i\frac{\partial}{\partial t}-H_0\right)
\left[\begin{array}{c}
\phi^s_{n,\frac{3}{2}}\\
\phi^s_{n,\frac{1}{2}}\\
\phi^s_{n,-\frac{1}{2}}\\
\phi^s_{n,-\frac{3}{2}}
\end{array}\right]
=\frac{g_0}{2}\sum_{n's'}M\left[\begin{array}{c}
\phi^s_{n,\frac{3}{2}}\\
\phi^s_{n,\frac{1}{2}}\\
\phi^s_{n,-\frac{1}{2}}\\
\phi^s_{n,-\frac{3}{2}}
\end{array}\right],
\label{eq:hf}
\end{eqnarray}
where $t=\omega\tilde{t}$ is the reduced time with $\tilde{t}$ the real time and the interaction matrix
\begin{eqnarray}
M=\left[\begin{array}{cccc}
\vert\phi^{s'}_{n',-\frac{3}{2}}\vert^2&
-\mathcal{M}_{1}&
\mathcal{M}_{2}&
-\mathcal{M}_{3}\\
-\mathcal{M}_{1}^*&
\vert\phi^{s'}_{n',-\frac{1}{2}}\vert^2&
-\mathcal{M}_{4}&
\mathcal{M}_{5}\\
\mathcal{M}_{2}^*&
-\mathcal{M}_{4}^*&
\vert\phi^{s'}_{n',\frac{1}{2}}\vert^2&
-\mathcal{M}_{6}\\
-\mathcal{M}_{3}^*&
\mathcal{M}_{5}^*&
-\mathcal{M}_{6}^*&
\vert\phi^{s'}_{n',\frac{3}{2}}\vert^2
\end{array}\right],
\end{eqnarray}
with
\begin{eqnarray}
\mathcal{M}_{1}&=&\phi^{s'*}_{n',-\frac{3}{2}}\phi^{s'}_{n',-\frac{1}{2}},
\mathcal{M}_{2}=\phi^{s'*}_{n',-\frac{3}{2}}\phi^{s'}_{n',\frac{1}{2}},
\nonumber\\
\mathcal{M}_{3}&=&\phi^{s'*}_{n',-\frac{3}{2}}\phi^{s'}_{n',\frac{3}{2}},
\mathcal{M}_{4}=\phi^{s'*}_{n',-\frac{1}{2}}\phi^{s'}_{n',\frac{1}{2}},
\nonumber\\
\mathcal{M}_{5}&=&\phi^{s'*}_{n',-\frac{1}{2}}\phi^{s'}_{n',\frac{3}{2}},
\mathcal{M}_{6}=\phi^{s'*}_{n',\frac{1}{2}}\phi^{s'}_{n',\frac{3}{2}}.
\end{eqnarray}
These HF equations govern the spinor dynamics of the spin $f=3/2$ Fermi systems. It accounts for the first-order contribution to the non-interaction energy. Thus it is valid at weak
interaction $k_Fa_0\ll1$. In the following, we choose $E_F=\hbar^2k_F^2/(2m)\sim\hbar\omega$ and $g_0=0.2$, thus $k_F\sim\sqrt{m\omega/\hbar}$ and $a_0\ll0.1\sqrt{\hbar/(m\omega)}$. So we have $k_Fa_0\ll 1$ and we can safely use HF theory in this regime.

\section{Two-Body oscillation}
\begin{figure}[h]
\includegraphics[scale=0.8]{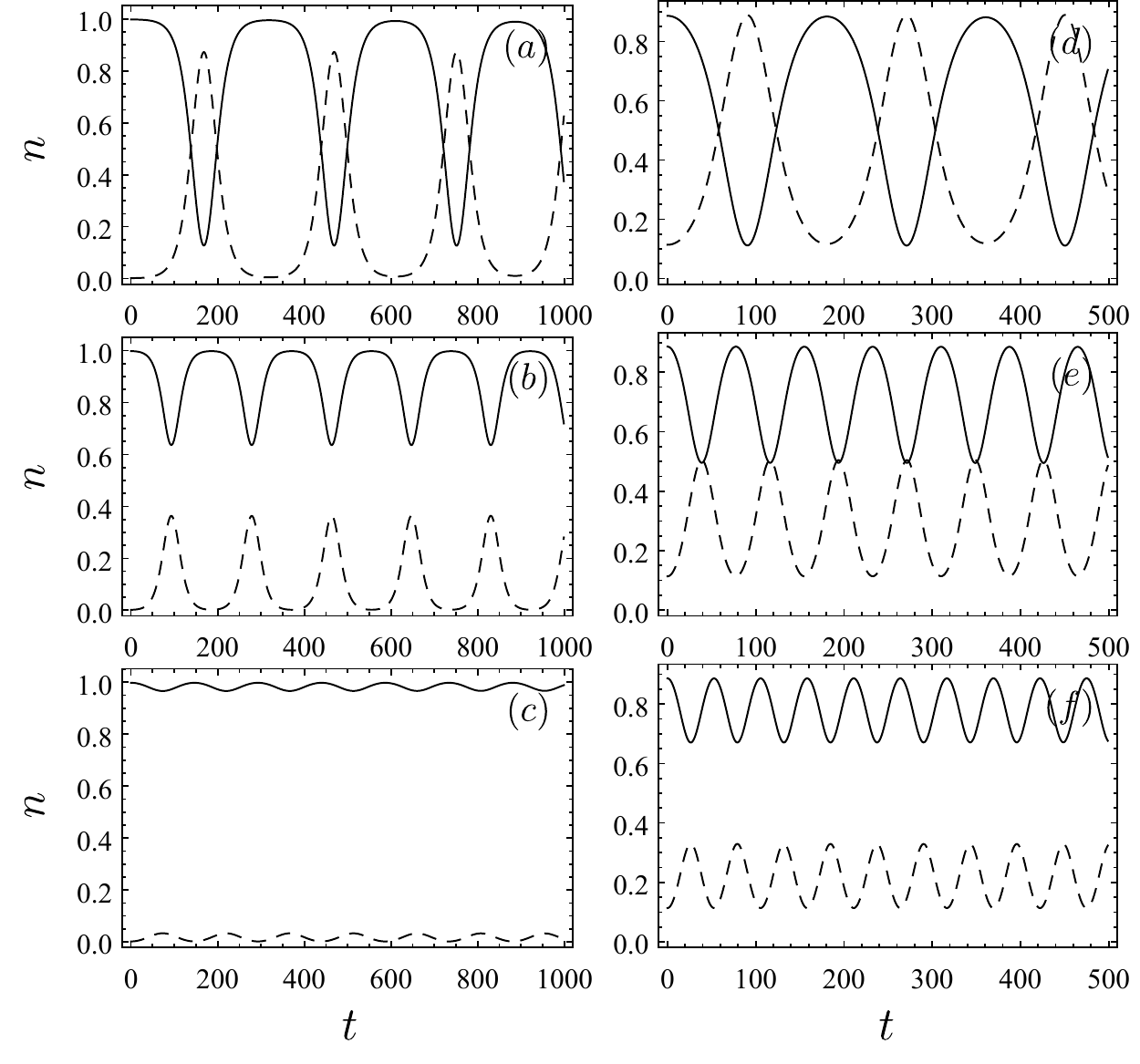}
 \centering
\caption{Two-body relative occupation number of spin state $m=\pm1/2$ (Solid) and $m=\pm3/2$ (Dashed) as a function of time for different quadratic Zeeman energy $q=0.005,0.025,0.04$ from $(a)$-$(c)$ for $\theta=0.05$ and from $(d)$-$(f)$ for $\theta=0.4$. The interaction strength is $g_0=0.2$.}
\label{fg:fig1}
\end{figure}

We first see how the two-body system evolves in this case. We note the two-body dynamics here can also be separated to center-of mass and relative motion, which results in the relative coordinate coupled-channel Schr\"odinger equations as studied in \cite{OHara}. At weak interaction $k_Fa_0\ll1$, our HF theory should give the same physics as this method. As mentioned in our previous section, the two atoms are initially prepared in a degenerate state ($n=1$) with spin $\up$ and $\dn$, with the wavefunction of the system
\begin{equation}
\Psi(\mathbf{x})=\frac{1}{\sqrt{2}}\left[
\psi_{1}^{\uparrow}(x_{1})\psi_{1}^{\downarrow}(x_{2})-
\psi_{1}^{\uparrow}(x_{2})\psi_{1}^{\downarrow}(x_{1})
\right],
\end{equation}
where $\psi_{1}^{\up/\dn}(x)$ is defined in Eq.~\ref{eq:singleparticlewf}. We get the dynamics by solving corresponding HF equations in Eq.~(\ref{eq:hf}). Similar like the experiment in \cite{Krauser2}, we choose two orthogonal state as our initial state, i.e., $\psi_1^\up(x)=\exp(-iS_x\theta)[0,1,0,0]^Tc_1(x)$, $\psi_1^\dn(x)=\exp(-iS_x\theta)[0,0,1,0]^Tc_1(x)$, where $c_{n=1}(x)$ is the $n$-th harmonic oscillator eigenstate, $S_x$ is the spin matrix in the $x$ direction and $\theta$ is the corresponding rotation angle. Experimentally, this is achieved by introducing a radio-frequency pulse to rotate the $m=\pm1/2$ polarized spins about the $x$-axis with angle $\theta$. We show the evolution of relative occupation number at different quadratic Zeeman energy for rotation angle $\theta=0.05$ and $0.4$ in Fig.~\ref{fg:fig1}. The interaction induces a Rabi oscillation between $m=\pm1/2$ and $m=\pm3/2$ states. The energy difference of these two sates is tuned by the quadratic magnetic energy $q$. So for increasing magnetic field, the mean relative occupation number of $m=\pm1/2$ state is increasing as illustrated in Fig.~\ref{fg:fig1}.

\begin{figure}[h]
 \centering
\includegraphics[scale=0.85]{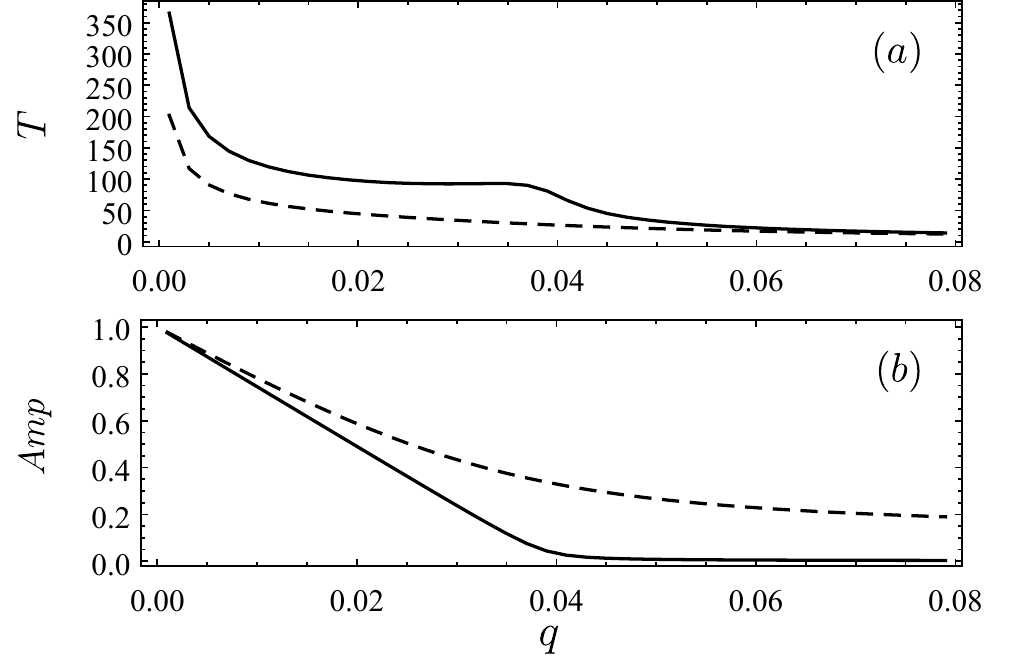}
\caption{Two-body oscillation period and relative amplitude for spin state $m=\pm1/2$ as a function of quadratic Zeeman energy $q$ at interaction strength $g_0=0.2$. The solid (dashed) line corresponds to rotation angle $\theta=0.05(0.4)$.}
\label{fg:fig2}
\end{figure}

We notice in Fig.~\ref{fg:fig1} from $(a)$-$(c)$ there seems a transition to the over oscillated regime, where the spins hold for a while before flipping to other spin state. For $\theta=0.4$ this transition is absent. Similar transition arises when we plot the oscillation periods and relative amplitude for different quadratic Zeeman energy $q$ in Fig.~\ref{fg:fig2}. We see there is a shoulder in oscillation period for rotation angle $\theta=0.05$, while for $\theta=0.4$ the period is monotonously decreasing for increasing magnetic field.

\begin{figure}[h]
 \centering
\includegraphics[scale=0.75]{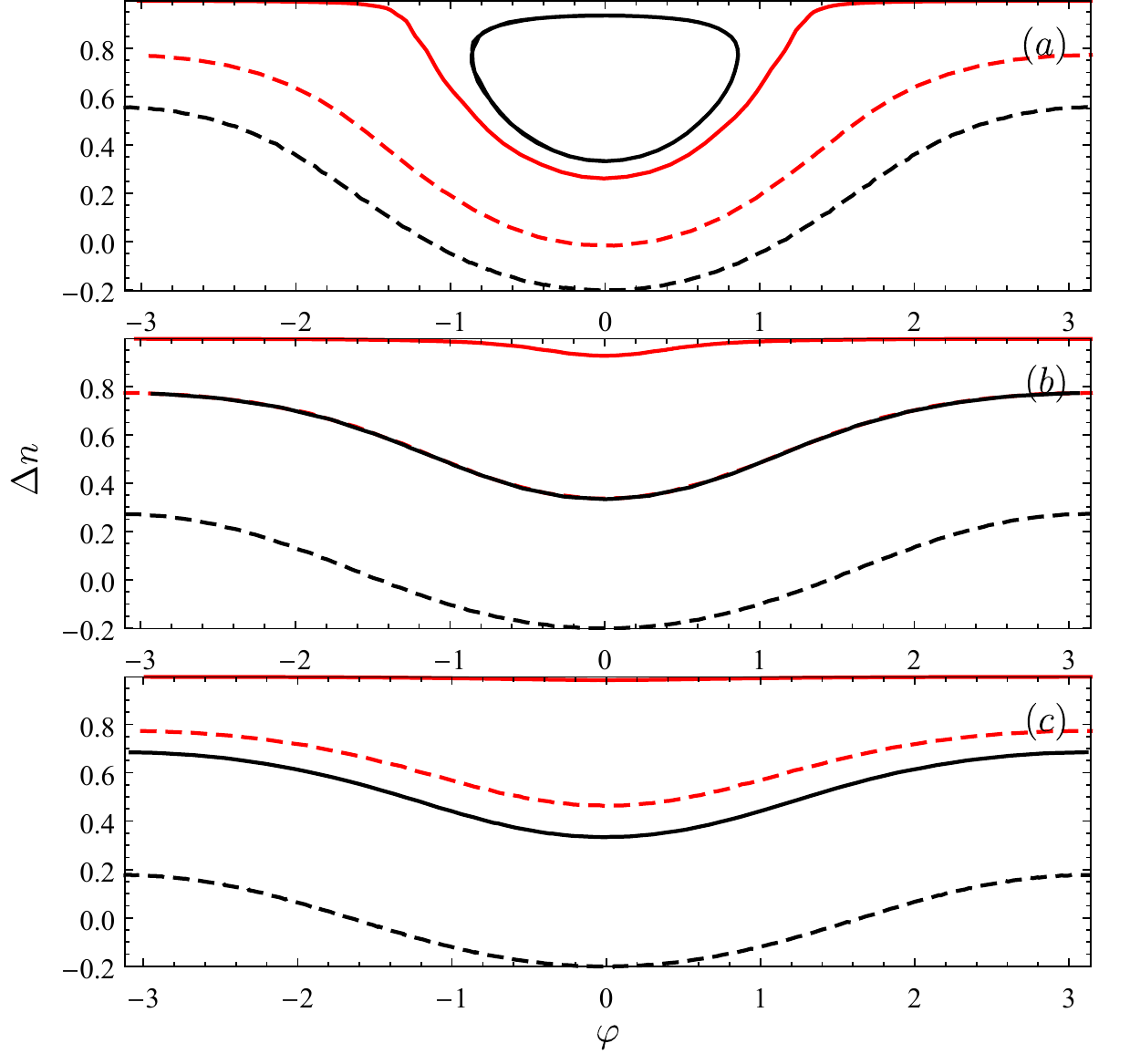}
\caption{Phase space trajectory for rotation angle $\theta=0.05$ (solid red) and $0.4$ (dashed red). The quadratic Zeeman energy $q=0.025, 0.04, 0.05$ from $(a)$-$(c)$ and the interaction strength $g_0=0.2$.}
\label{fg:fig3}
\end{figure}

Similar behaviors have been observed in spinor Bose-Einstein condensates, where the transition is found associated with the phase space topology from periodic phase to running phase, which is similar like a classical non-rigid pendulum \cite{Zhang}. Whether the transition here corresponds to the same mechanism? To answer this, we first define the relative phase and population difference between $m=\pm3/2$ and $\pm1/2$ state as $\varphi(x)=\arg(\phi^{\up*}_{1,\frac{3}{2}}\phi^{\up*}_{1,-\frac{3}{2}}\phi^{\up}_{1,\frac{1}{2}}\phi^{\up}_{1,-\frac{1}{2}})$, $\Delta n=n_{1/2}+n_{-1/2}-n_{3/2}-n_{-3/2}$. We find the local relative phase is almost the same for the majority of the fermions, so we will use the relative phase $\varphi=\varphi(x=0)$ hereafter. The evolution of the system forms a contour in the $\Delta n-\varphi$ plane, which is shown in Fig.~\ref{fg:fig3}. The solid and dashed red lines correspond to rotation angle $\theta=0.05$ and $0.4$. For comparison, we also plot the case with initial relative phase $\varphi=0$ and population difference $\Delta n=0.33$ and $-0.2$ for solid and dashed black lines (see Fig.~\ref{fg:fig3}). The quadratic Zeeman energy $q=0.025, 0.04, 0.05$ from $(a)$-$(c)$. For the solid black line, there is a transition from periodic phase (where the contour forms a closed circle) in Fig.~\ref{fg:fig3}$(a)$ to running phase in Fig.~\ref{fg:fig3}$(c)$, and the oscillation period diverges at the resonance point \cite{Kawaguchi,Zhang}. For the initial condition in dashed black line, we can not see such resonance for increasing magnetic field.

For the initial conditions considered here (red lines), we see for both rotation angles the contours do not form a closed circle, thus the relative phases are all running monotonously. So the shoulder in the oscillation period in Fig.~\ref{fg:fig2} does not correspond to the resonance as in \cite{Zhang}. In fact, it is not a resonance here. For $\theta=0.4$ we can not see any shoulders in the oscillation period as in Fig.~\ref{fg:fig2}.

\begin{figure}[h]
 \centering
\includegraphics[scale=0.65]{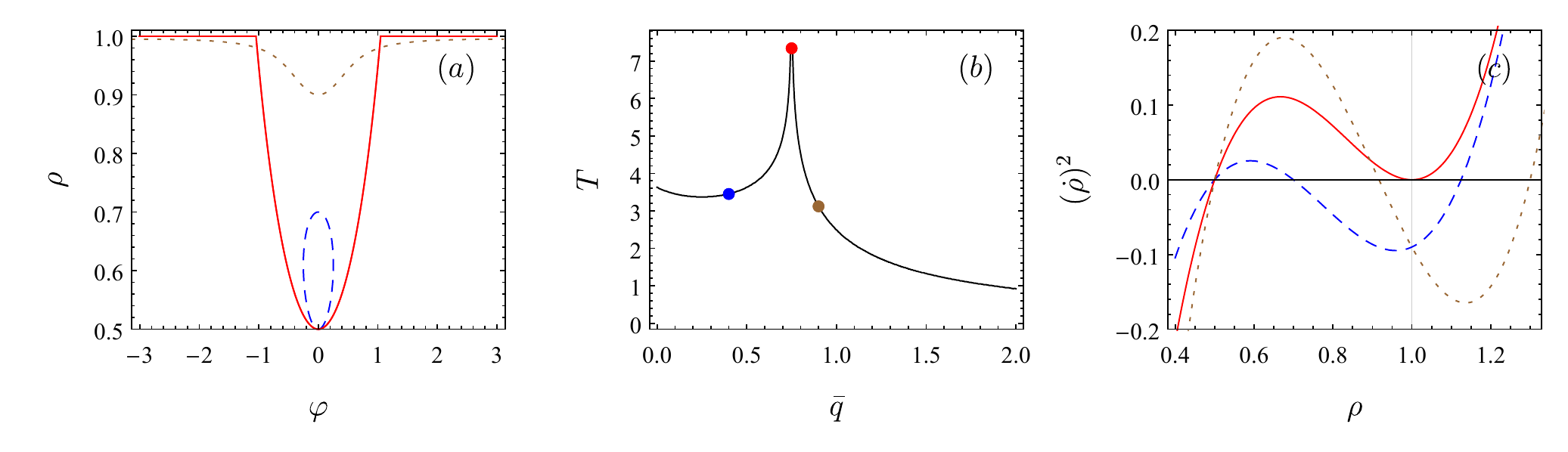}
\caption{Initial state with $\rho_0=0.5, \varphi_0=0$. The blue/red/brown color stands for reduced quadratic Zeeman energy $\bar{q}=0.6/0.75/0.9$. (a) The phase space trajectory shows a transition from periodic to running phase, which is consistent with the oscillation period plot in (b), where the period $T$ diverges at $\bar{q}=0.75$. At the transition point the two largest solutions to $(\dot{\rho})^2=0$ touch each other with $x_2=x_3=1$, which is shown in (c).}
\label{fg:fig4}
\end{figure}

To further understand this, we consider a simplified model with wavefunction for $m=3/2,1/2,-1/2,-3/2$ states as $c_1,c_2,c_3,c_4$. The mean field energy of the system is written as
\begin{eqnarray}
E=&\frac{9}{4}\bar{q}\left(|c_1|^2+|c_4|^2\right)+\frac{1}{4}\bar{q}\left(|c_2|^2+|c_3|^2\right)\nonumber\\
&+|c_1|^2|c_4|^2+|c_2|^2|c_3|^2-c_1c_2^*c_3^*c_4-c_1^*c_2c_3c_4^*,
\label{eq:mfEn}
\end{eqnarray}
with the dynamics given by $i\partial_t\Psi=H\Psi$ where
\begin{eqnarray}
H=\left[\begin{array}{cccc}
\frac{9}{4}\bar{q}+\vert c_4\vert^2&-c_4^*c_3&0&0\\
-c_3^*c_4&\frac{1}{4}\bar{q}+\vert c_3\vert^2&0&0\\
0&0&\frac{1}{4}\bar{q}+\vert c_2\vert^2&-c_2^*c_1\\
0&0&-c_1^*c_2&\frac{9}{4}\bar{q}+\vert c_1\vert^2
\end{array}\right],
\end{eqnarray}
and we have used the reduced quadratic Zeeman energy $\bar{q}$ and time $\bar{t}$. Due to the symmetry of $\pm3/2(1/2)$ states, we set $c_1(c_2)=c_4(c_3)$ and the Schr\"{o}dinger equation is
\begin{eqnarray}
i\partial_{\bar{t}}
\left[\begin{array}{c}
c_1\\c_2
\end{array}\right]
=\left[\begin{array}{cc}
\frac{9}{4}\bar{q}+\vert c_1\vert^2&-c_1^*c_2\\
-c_2^*c_1&\frac{1}{4}\bar{q}+\vert c_2\vert^2
\end{array}\right]
\left[\begin{array}{c}
c_1\\c_2
\end{array}\right].
\label{eq:toy}
\end{eqnarray}
Since the wavefunction is independent of $x$, we can write $c_1=\sqrt{\rho_1}e^{i\theta_1}$, $c_2=\sqrt{\rho_2}e^{i\theta_2}$. The relative population and phase is defined as $\rho=\rho_2-\rho_1$ and $\varphi=2(\theta_2-\theta_1)$. From Eq. (\ref{eq:toy}) the relative phase and population obeys
\begin{eqnarray}
\dot{\rho}=(1-\rho^2)\sin\varphi,\label{eq:rho}\\
\dot{\varphi}=4q-2\rho(1+\cos\varphi),
\end{eqnarray}
which is also consistent with the relation $\dot{\varphi}=-2\partial E/\partial\rho$, $\dot{\rho}=2\partial E/\partial\varphi$ where the mean-field energy of Eq. (\ref{eq:mfEn}) $E=[1+5q-4q\rho+\rho^2-(1-\rho^2)\cos\varphi]/2$. The contour plot of the energy in Fig. \ref{fg:fig4}a shows a transition from periodic to running phase for an initial state $\rho_0=0.5, \varphi_0=0$. At the transition point the oscillation period diverges as in Fig. \ref{fg:fig4}b. The plot of $(\dot{\rho})^2=(1-\rho^2)^2-(1+5q-4q\rho+\rho^2-2E)^2$ is shown in Fig. \ref{fg:fig4}c with zero crossing at $x_1,x_2,x_3$ ($x_1<x_2<x_3$). Since $(\dot{\rho})^2>0$, the relative population oscillates between $x_1$ and $x_2$. The critical point happens when $x_2=x_3=1$, which gives $\bar{q}_c=2E-2=(1+\rho_0)/2=0.75$, as is also shown at the red point in Fig. \ref{fg:fig4}b.

\begin{figure}[h]
 \centering
\includegraphics[scale=0.55]{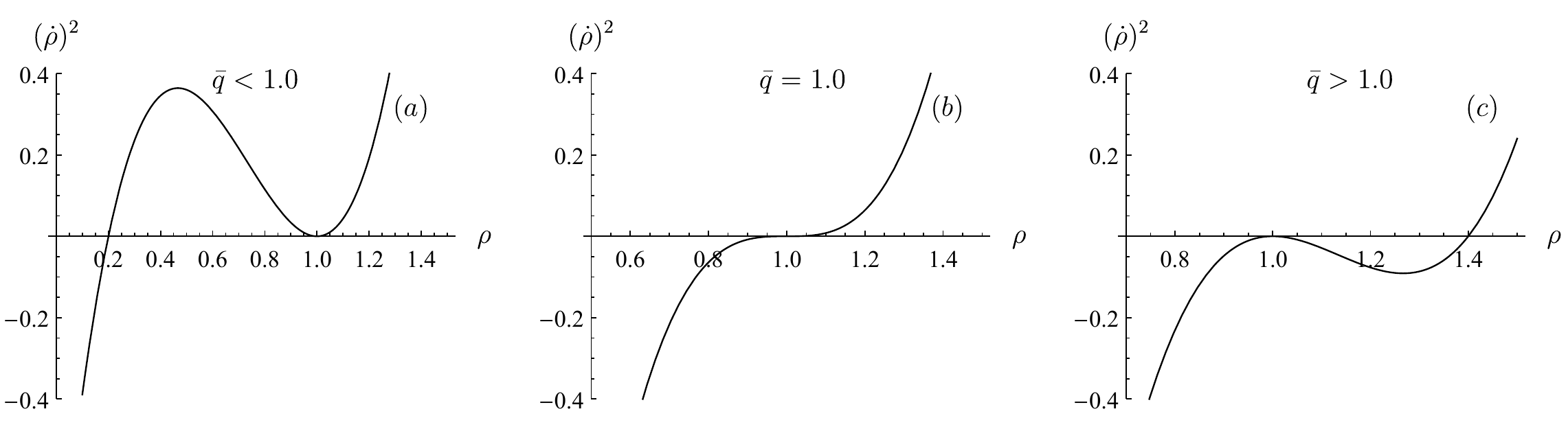}
\caption{Plot of $(\dot{\rho})^2$ for different reduced quadratic Zeeman energy $\bar{q}$ at the limit of zero rotation angle. (a) For $\bar{q}<1$ the system is in a transition phase as the red curve in Fig. \ref{fg:fig4}c. The critical point happens at (b) $\bar{q}=1$ where there is only one solutions to $(\dot{\rho})^2=0$ and the oscillation period diverges. Above this point the relative number $\rho$ is locked to $1$ as in (c).}
\label{fg:fig5}
\end{figure}

\begin{figure}[h]
 \centering
\includegraphics[scale=0.65]{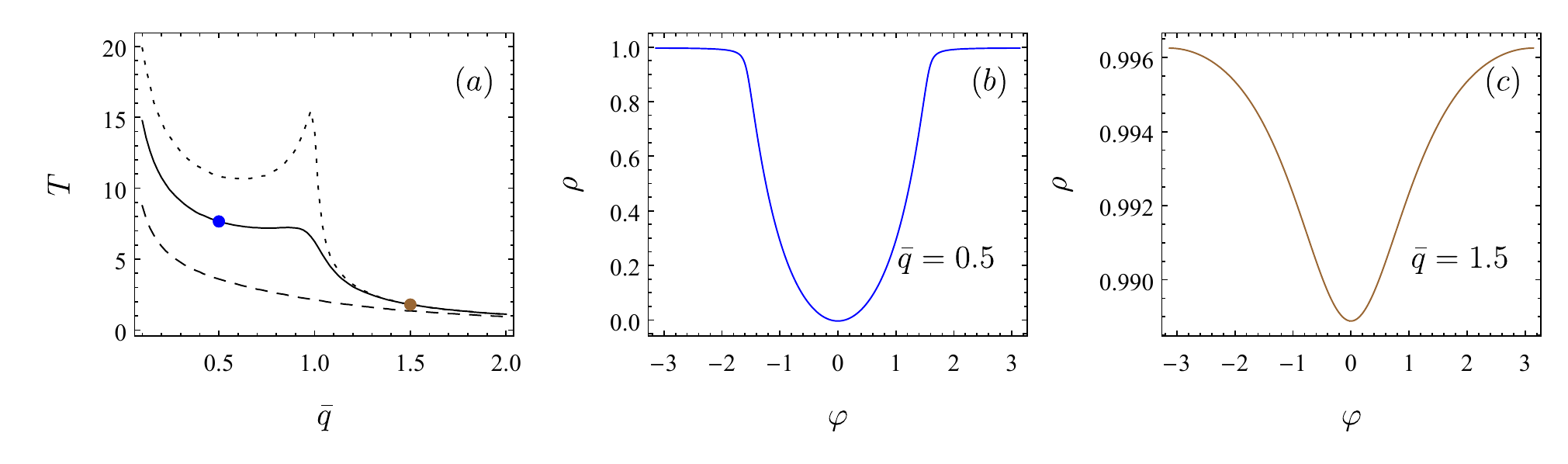}
\caption{(a) Oscillation period as a function of reduced quadratic Zeeman energy $\bar{q}$. The dotted/solid/dashed curve corresponds to rotation angle $\theta=0.01/0.05/0.4$. (b) and (c) Phase space trajectory for rotation angle $\theta=0.05$ for reduced quadratic Zeeman energy $\bar{q}=0.5(a)$ and $1.5(b)$.}
\label{fg:fig6}
\end{figure}

Typically for a initial state with finite relative population $\rho$ and zero phase $\varphi$, there is always a transition from periodic to running phase (as is shown in Fig. \ref{fg:fig4}). How about the experimentally initial state with rotation angle $\theta$, i.e., $c_1(\bar{t}=0)=\sqrt{3}(-e^{i\theta/2}+e^{-3i\theta/2})/4$ and $c_2(\bar{t}=0)=(e^{i\theta/2}+3e^{-3i\theta/2})/4$? Before answering this, let's first see the limiting case with $\theta\to0$, which corresponds to initial state $\rho_0=1,\varphi_0=0$ and $(\dot{\rho})^2=8q(1-\rho)^2(1+\rho-2q)$. We plot $(\dot{\rho})^2$ in Fig. \ref{fg:fig5}. At $\bar{q}<1$ we have $x_2=x_3=1$ and the system is in the transition phase where $\rho$ oscillates between $2q-1$ to $1$. When $\bar{q}>1$, the relative population is locked to $1$. The critical point happens at $\bar{q}=1.0$ and the system never goes into the periodic phase.

The zero rotation angle can be thought of limiting cases of our problem. We show the oscillation period for $\theta=0.05(0.4)$ in Fig. \ref{fg:fig6} as the solid (dashed) line. The oscillation period peak at $\bar{q}=1.0$ becomes a shoulder because there is not a transition for finite $\theta$. The phase space trajectory in Fig.~\ref{fg:fig6}b indicates the system is always in a running phase regime. Way below and above the shoulder position, the phase space is close to a resonance (Fig.~\ref{fg:fig6}b and Zeeman (Fig.~\ref{fg:fig6}c) regime as in Fig.~\ref{fg:fig3} respectively. Such behaviour reminds us the case of magnetically tuned ferromagnetic spinor $^{87}{\rm Rb}$ condensates studied in \cite{Zhang}. However, the oscillation period shoulder in the latter case corresponds to the resonance from periodic phase to running phase in the limit of zero magnetization. At last, we note the similar rotation angle dependent behaviour should also appear in spinor Bose-Einstein condensates.

\section{Many-Body oscillation}
\begin{figure}[h]
 \centering
\includegraphics[scale=0.8]{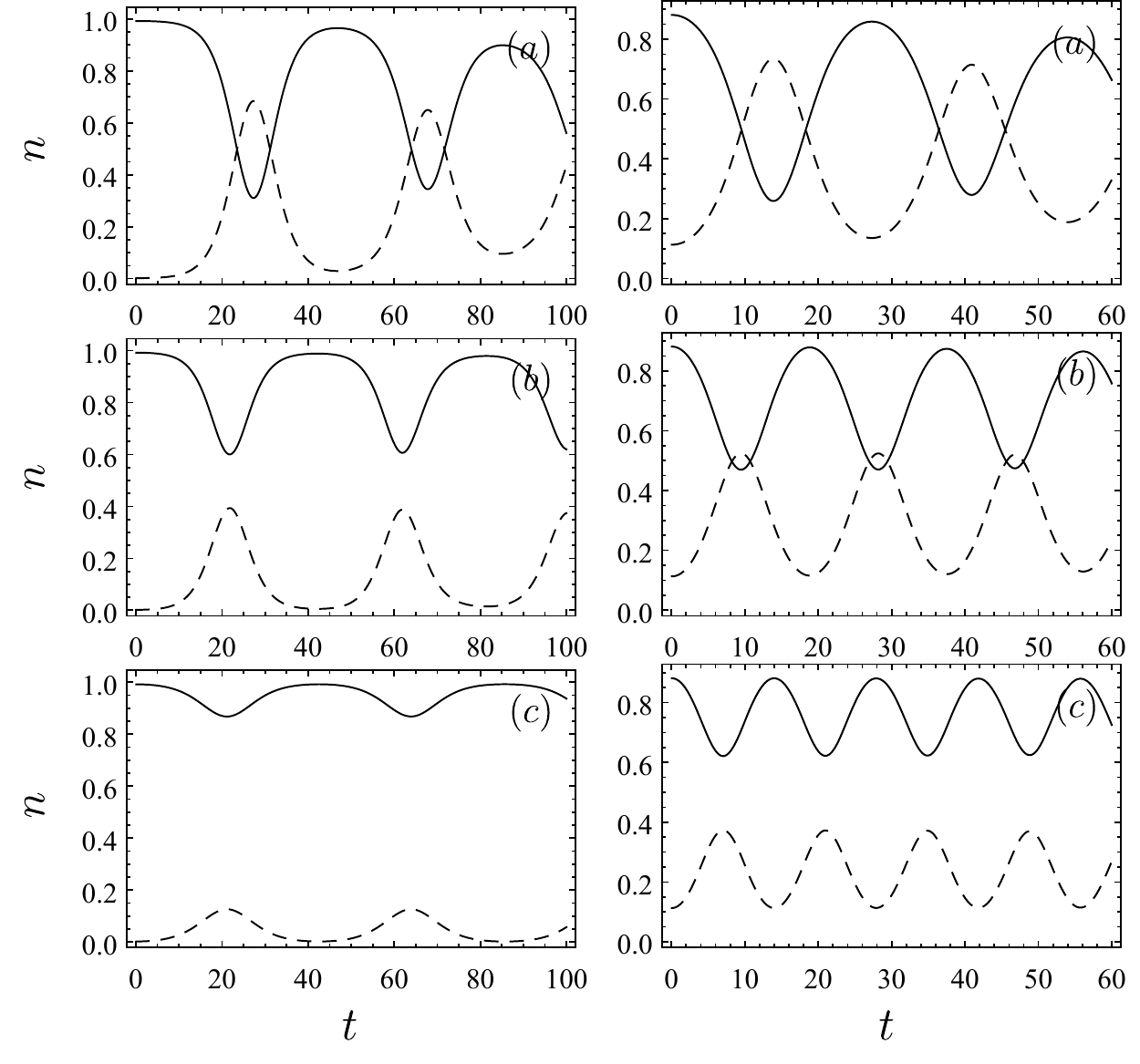}
\caption{Many-body relative occupation number of spin state $m=\pm1/2$ (Solid) and $m=\pm3/2$ (Dashed) as a function of time for different quadratic Zeeman energy $q=0.05,0.1,0.15$ from $(a)$-$(c)$ for $\theta=0.05$ and from $(d)$-$(f)$ for $\theta=0.4$. The interaction strength is $g_0=0.2$ and particle number $N=40$.}
\label{fg:fig7}
\end{figure}

The collective oscillation of fermions is different from Bose gases due to the Pauli exclusion principle, which prohibits their condensation into one single-particle state. Fermi gases in this case will have multi-oscillation modes, which lead the collective oscillation to decay. We now go to many-body physics to see whether the collective oscillation and shoulder still preserve in a Fermi sea. Like our two-body case, the initial single-particle wavefunction is taken as  $\psi_n^\up(x)=\exp(-iS_x\theta)[0,1,0,0]^Tc_n(x)$, $\psi_n^\dn(x)=\exp(-iS_x\theta)[0,0,1,0]^Tc_n(x)$, where $\theta=0.05$ and $0.4$ respectively. We plot in Fig.~\ref{fg:fig7} the relative population as a function of time for $N=40$ particles. We see the collective oscillation preserves well even in the many-body Fermi sea. In the weak interaction we are considering, the collective oscillation can be considered as a superposition of different single-particle oscillations with small frequency shifts. When go to long time dynamics, there will be dephasing and multi-oscillation modes will appear.

\begin{figure}[h]
 \centering
\includegraphics[scale=0.8]{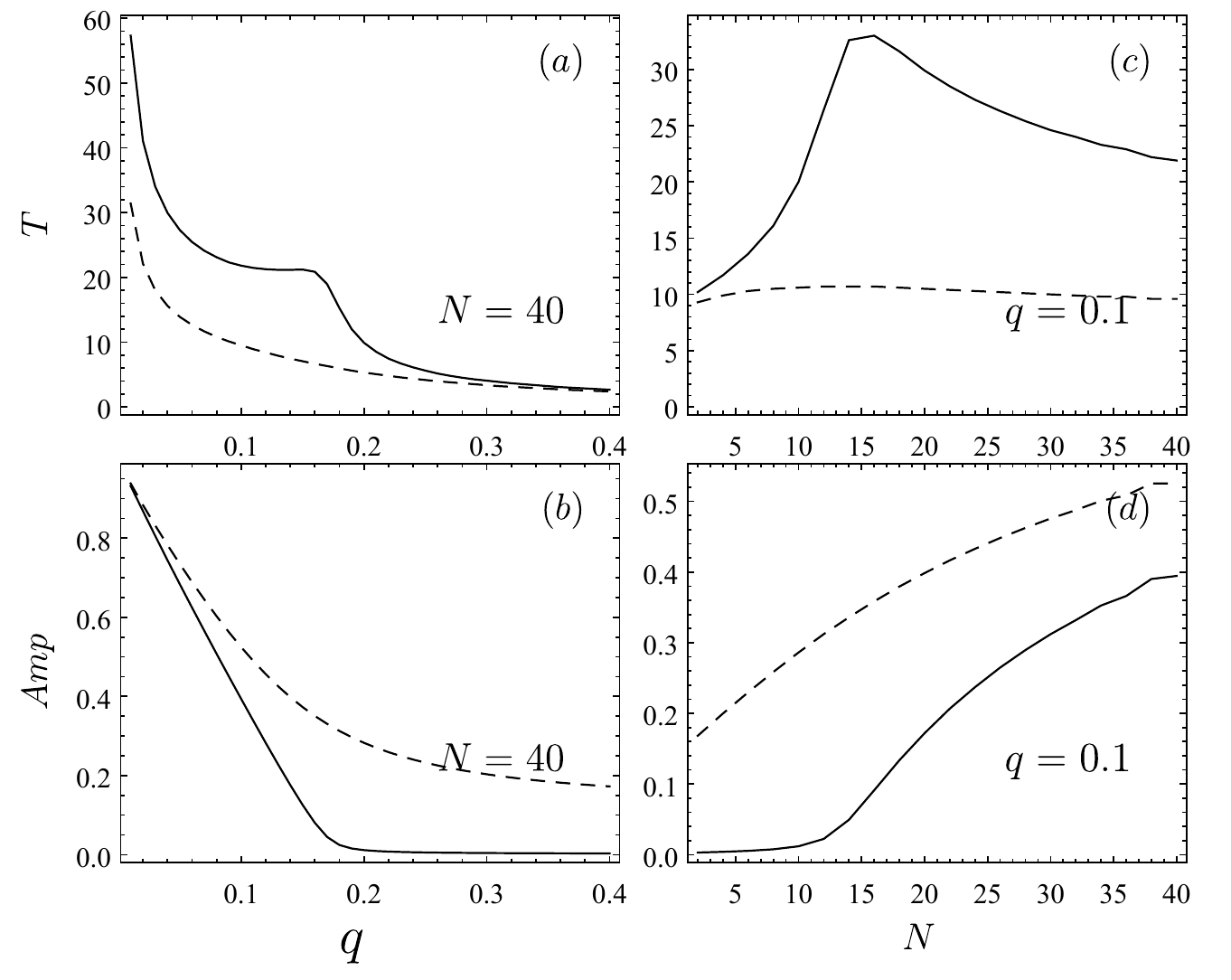}
\caption{(a-b) Many-body oscillation period and relative amplitude for spin state $m=\pm1/2$ as a function of quadratic Zeeman energy $q$ at interaction strength $g_0=0.2$. The solid (dashed) line corresponds to rotation angle $\theta=0.05(0.4)$. The particle number  is $N=40$. (c-d) Many-body oscillation period and relative amplitude for spin state $m=\pm1/2$ as a function of particle number $N$ at interaction strength $g_0=0.2$. The quadratic Zeeman energy is $q=0.1$.}
\label{fg:fig8}
\end{figure}

The oscillation shoulder, which is a remnant of the critical point in the two-body phase space, still survives in the many-body case. It appears at small rotation angles for $N=40$ particles in Fig.~\ref{fg:fig8}a. We also show the oscillation period and amplitude as a function of particle number for quadratic magnetic field $q=0.1$. It shows by increase the particle number, the system will evolve from above to below the shoulder, resulting in a maximum in the oscillation period as shown in Fig. \ref{fg:fig8}c. This two- to many-body evolution could be tested in the future experiments.

\section{Summary}
In this paper, we have developed a Hartree-Fock theory to study the collective spin-mixing dynamics of spin-3/2 fermions. This method in principle can be generalized to arbitrary high-spin fermions. We find the rotation angle of the initial state can give rise to a shoulder in the oscillation period. Different from previous studies, where the shoulder is found connected to the resonance from periodic to running phase, the system is always in a running phase in the two-body phase space. This shoulder survives even in the many-body oscillations, which could be tested in the experiments. We also give a picture how these dynamics evolves from two- to many-body. These studies complement our understanding of collective dynamics of large-spin Fermi systems.

Even though there are additional orbital states in spinor fermions, compared to the Bose-Einstein condensates, their dynamics is highly suppressed due to the existence of the Fermi sea. It would be interesting to study how the interorbital changing influences the dynamics as the temperature is increasing. Also we note our HF equations here only work for weak interaction and do not take into account the collisional effect. The latter effect should be proportional to $g_0^2$ and gives rise to damping and relaxation in the system \cite{Ebling}.

\section*{ACKNOWLEDGMENTS}
This research is supported by the National Key Basic Research Program of China (No. 2013CB922002) and the National Natural Science Foundation of China (No. 11574028, 11504021). J.X. is also supported by CPSF (No. 2015M580043, 2016T90032), and FRFCU (No. FRF-TP-15-040A1).

\end{document}